\documentclass[journal]{IEEEtran}

\IEEEoverridecommandlockouts
\usepackage{acronym}
\usepackage{amsmath,amssymb,amsfonts}
\usepackage{algorithmic}
\usepackage{textcomp}
\usepackage{xcolor}
\usepackage[pdftex]{graphicx}

\usepackage[utf8]{inputenc}
\usepackage[T1]{fontenc}
\usepackage{microtype}
\usepackage{mathtools,gensymb,eurosym,url,soul} % provides amsmath, generic symbols, URL management and strikeout
\usepackage{subfig}
\usepackage{graphicx}
\usepackage{xcolor}%[usenames,dvipsnames,svgnames,table]{xcolor}
\usepackage[citestyle=numeric-comp,bibstyle=ieee,sorting=none,firstinits,maxbibnames=5,backend=biber]{biblatex}
\usepackage{hyperref}

\hypersetup{colorlinks=true, breaklinks=true, %pagebackref=true,
  urlcolor=blue, linkcolor=blue,anchorcolor=blue,citecolor=blue,
  pdfpagemode = UseNone, %FullScreen, %UseThumbs, %UseOutlines,
  pdfauthor = {},
  pdftitle = {},
  pdfsubject = {},
  pdfkeywords = {}
}

\graphicspath{{./img/}{./scripts/}}
\DeclareGraphicsExtensions{.pdf,.jpeg,.png,.jpg}

\DeclareSourcemap{
	\maps[datatype=bibtex]{
		\map[overwrite=true]{
			\step[fieldset=doi, null]
		}
	}
}

\newcommand{\sd}{$\Sigma\Delta$ }
\newcommand{\vn}{von Neumann }

\acrodefplural{RAM}[RAMs]{Random Access Memories}
\acrodefplural{RRAM}[R-RAMs]{Resistive Random Access Memories}
\acrodefplural{STT-MRAM}[STT-MRAMs]{Spin-Transfer Torque Magnetic Random Access Memories}
\acrodef{ADC}[ADC]{Analog to Digital Converter}
\acrodef{ADEX}[AdExpIF]{Adaptive-Exponential Integrate and Fire}
\acrodef{AER}[AER]{Address-Event Representation}
\acrodef{AEX}[AEX]{AER EXtension board}
\acrodef{AE}[AE]{Address-Event}
\acrodef{AFM}[AFM]{Atomic Force Microscope}
\acrodef{AGC}[AGC]{Automatic Gain Control}
\acrodef{AMDA}[AMDA]{AER Motherboard with D/A converters}
\acrodef{ANN}[ANN]{Attractor Neural Network}
\acrodef{API}[API]{Application Programming Interface}
\acrodef{ARM}[ARM]{Advanced RISC Machine}
\acrodef{ASIC}[ASIC]{Application Specific Integrated Circuit}
\acrodef{BCM}[BMC]{Bienenstock-Cooper-Munro}
\acrodef{BD}[BD]{Bundled Data}
\acrodef{BEOL}[BEOL]{Back-end of Line}
\acrodef{BG}[BG]{Bias Generator}
\acrodef{BMI}[BMI]{Brain-Machince Interface}
\acrodef{BPTT}[BPTT]{Back-Propagation Through Time}
\acrodef{CAD}[CAD]{Computer Aided Design}
\acrodef{CAM}[CAM]{Content Addressable Memory}
\acrodef{CAVIAR}[CAVIAR]{Convolution AER Vision Architecture for Real-Time}
\acrodef{CFC}[CFC]{Current to Frequency Converter}
\acrodef{CCN}[CCN]{Cooperative and Competitive Network}
\acrodef{CDR}[CDR]{Clock-Data Recovery}
\acrodef{CFC}[CFC]{Current to Frequency Converter}
\acrodef{CHP}[CHP]{Communicating Hardware Processes}
\acrodef{CNN}[CCN]{Convolutional Neural Network}
\acrodef{CMIM}[CMIM]{Metal-insulator-metal Capacitor}
\acrodef{CML}[CML]{Current Mode Logic}
\acrodef{CMOL}[CMOL]{``Hybrid CMOS nanoelectronic circuits''}
\acrodef{CMOS}[CMOS]{Complementary Metal-Oxide-Semiconductor}
\acrodef{CNN}[CCN]{Convolutional Neural Network}
\acrodef{COTS}[COTS]{Commercial Off-The-Shelf}
\acrodef{CPG}[CPG]{Central Pattern Generator}
\acrodef{CPLD}[CPLD]{Complex Programmable Logic Device}
\acrodef{CPU}[CPU]{Central Processing Unit}
\acrodef{CSM}[CSM]{Cortical State Machine}
\acrodef{CSP}[CSP]{Constraint Satisfaction Problem}
\acrodef{CV}[CV]{Coefficient of Variation}
\acrodef{DAC}[DAC]{Digital to Analog Converter}
\acrodef{DAS}[DAS]{Dynamic Auditory Sensor}
\acrodef{DAVIS}[DAVIS]{Dynamic and Active Pixel Vision Sensor}
\acrodef{DBN}[DBN]{Deep Belief Network}
\acrodef{DFA}[DFA]{Deterministic Finite Automaton}
\acrodef{DI}[DI]{delay insensitive}
\acrodef{DMA}[DMA]{Direct Memory Access}
\acrodef{DNF}[DNF]{Dynamic Neural Field}
\acrodef{DNN}[DNN]{Deep Neural Network}
\acrodef{DOF}[DOF]{Degrees of Freedom}
\acrodef{DPE}[DPE]{Dynamic Parameter Estimation}
\acrodef{DPI}[DPI]{Differential Pair Integrator}
\acrodef{DR-RZ}[DR-RZ]{Dual-Rail Return-to-Zero}
\acrodef{DRAM}[DRAM]{Dynamic Random Access Memory}
\acrodef{DR}[DR]{Dual Rail}
\acrodef{DSP}[DSP]{Digital Signal Processor}
\acrodef{DVS}[DVS]{Dynamic Vision Sensor}
\acrodef{EBL}[EBL]{Electron Beam Lithography}
\acrodef{EDVAC}[EDVAC]{Electronic Discrete Variable Automatic Computer}
\acrodef{EIN}[EIN]{Excitatory-Inhibitory Network}
\acrodef{EM}[EM]{Expectation Maximization}
\acrodef{EPSC}[EPSC]{Excitatory Post-Synaptic Current}
\acrodef{EPSP}[EPSP]{Excitatory Post-Synaptic Potential}
\acrodef{ESN}[ESN]{Echo-State Network}
\acrodef{FDSOI}[FD-SOI]{Fully-Depleted Silicon on Insulator}
\acrodef{FET}[FET]{Field-Effect Transistor}
\acrodef{FFT}[FFT]{Fast Fourier Transform}
\acrodef{FI}[F-I]{Frequency-Current}
\acrodef{FPGA}[FPGA]{Field Programmable Gate Array}
\acrodef{FSA}[FSA]{Finite State Automaton}
\acrodef{FSM}[FSM]{Finite State Machine}
\acrodef{GOPS}[GOPS]{Giga-Operations per Second}
\acrodef{GPU}[GPU]{Graphical Processing Unit}
\acrodef{GUI}[GUI]{Graphical User Interface}
\acrodef{HAL}[HAL]{Hardware Abstraction Layer}
\acrodef{HH}[H\&H]{Hodgkin \& Huxley}
\acrodef{HMM}[HMM]{Hidden Markov Model}
\acrodef{HRS}[HRS]{High-Resistive State}
\acrodef{HR}[HR]{Human Readable}
\acrodef{HSE}[HSE]{Handshaking Expansion}
\acrodef{HW}[HW]{Hardware}
\acrodef{ICT}[ICT]{Information and Communication Technology}
\acrodef{IC}[IC]{Integrated Circuit}
\acrodef{IF2DWTA}[IF2DWTA]{Integrate \& Fire 2--Dimensional WTA}
\acrodef{IFSLWTA}[IFSLWTA]{Integrate \& Fire Stop Learning WTA}
\acrodef{IF}[I\&F]{Integrate-and-Fire}
\acrodef{IMU}[IMU]{Inertial Measurement Unit}
\acrodef{INCF}[INCF]{International Neuroinformatics Coordinating Facility}
\acrodef{INI}[INI]{Institute of Neuroinformatics}
\acrodef{IO}[I/O]{Input/Output}
\acrodef{IoT}[IoT]{Internet of Things}
\acrodef{IPSC}[IPSC]{Inhibitory Post-Synaptic Current}
\acrodef{IPSP}[IPSP]{Inhibitory Post-Synaptic Potential}
\acrodef{IP}[IP]{Intellectual Property}
\acrodef{ISI}[ISI]{Inter-Spike Interval}
\acrodef{IoT}[IoT]{Internet of Things}
\acrodef{JFLAP}[JFLAP]{Java - Formal Languages and Automata Package}
\acrodef{KCL}[KCL]{Kirchiff's current law}
\acrodef{LEDR}[LEDR]{Level-Encoded Dual-Rail}
\acrodef{LFP}[LFP]{Local Field Potential}
\acrodef{LIF}[LIF]{Leaky Integrate and Fire}
\acrodef{LLC}[LLC]{Low Leakage Cell}
\acrodef{LNA}[LNA]{Low-Noise Amplifier}
\acrodef{LPF}[LPF]{Low-Pass Filter}
\acrodef{LRS}[LRS]{Low-Resistive State}
\acrodef{LSM}[LSM]{Liquid State Machine}
\acrodef{LTD}[LTD]{Long Term Depression}
\acrodef{LTI}[LTI]{Linear Time-Invariant}
\acrodef{LTP}[LTP]{Long Term Potentiation}
\acrodef{LTU}[LTU]{Linear Threshold Unit}
\acrodef{LUT}[LUT]{Look-Up Table}
\acrodef{LVDS}[LVDS]{Low Voltage Differential Signaling}
\acrodef{MCMC}[MCMC]{Markov-Chain Monte Carlo}
\acrodef{MEMS}[MEMS]{Micro Electro Mechanical System}
\acrodef{MIM}[MIM]{Metal Insulator Metal}
\acrodef{MOSCAP}[MOSCAP]{Metal Oxide Semiconductor Capacitor}
\acrodef{MOSFET}[MOSFET]{Metal Oxide Semiconductor Field-Effect Transistor}
\acrodef{MOS}[MOS]{Metal Oxide Semiconductor}
\acrodef{MRI}[MRI]{Magnetic Resonance Imaging}
\acrodef{NDFSM}[NDFSM]{Non-deterministic Finite State Machine}
\acrodef{ND}[ND]{Noise-Driven}
\acrodef{NEF}[NEF]{Neural Engineering Framework}
\acrodef{NHML}[NHML]{Neuromorphic Hardware Mark-up Language}
\acrodef{NIL}[NIL]{Nano-Imprint Lithography}
\acrodef{NMDA}[NMDA]{N-Methyl-D-Aspartate}
\acrodef{NME}[NE]{Neuromorphic Engineering}
\acrodef{NRZ}[NRZ]{Non-Return-to-Zero}
\acrodef{NSM}[NSM]{Neural State Machine}
\acrodef{OTA}[OTA]{Operational Transconductance Amplifier}
\acrodef{PA}[PA]{Probabilistic Automata}
\acrodef{PCB}[PCB]{Printed Circuit Board}
\acrodef{PCHB}[PCHB]{Pre-Charge Half-Buffer}
\acrodef{PE}[PE]{Phase Encoding}
\acrodef{PCM}[PCM]{Phase Change Memory}
\acrodef{PFM}[PFM]{Pulse Frequency Modulation}
\acrodef{PR}[PR]{Production Rule}
\acrodef{PSC}[PSC]{Post-Synaptic Current}
\acrodef{PSP}[PSP]{Post-Synaptic Potential}
\acrodef{PSTH}[PSTH]{Peri-Stimulus Time Histogram}
\acrodef{QDI}[QDI]{Quasi Delay Insensitive}
\acrodef{RAM}[RAM]{Random Access Memory}
\acrodef{RELU}[ReLu]{Rectified Linear Unit}
\acrodef{RLS}[RLS]{Recursive Least-Squares}
\acrodef{RMSE}[RMSE]{Root Mean Squared-Error}
\acrodef{RMS}[RMS]{Root Mean Squared}
\acrodef{RNN}[RNN]{Recurrent Neural Network}
\acrodef{ROLLS}[ROLLS]{Reconfigurable On-Line Learning Spiking}
\acrodef{RRAM}[R-RAM]{Resistive Random Access Memory}
\acrodef{SAC}[SAC]{Selective Attention Chip}
\acrodef{SCX}[SCX]{Silicon CorteX}
\acrodef{SD}[SD]{Signal-Driven}
\acrodef{SDR}[SDR]{Signal to Distortion Ratio}
\acrodef{SEM}[SEM]{Spike-based Expectation Maximization}
\acrodef{SLAM}[SLAM]{Simultaneous Localization and Mapping}
\acrodef{SNN}[SNN]{Spiking Neural Network}
\acrodef{SRM}[SRM]{Spike Response Model}
\acrodef{SOC}[SOC]{System-On-Chip}
\acrodef{SOI}[SOI]{Silicon on Insulator}
\acrodef{SRAM}[SRAM]{Static Random Access Memory}
\acrodef{STDP}[STDP]{Spike-Timing Dependent Plasticity}
\acrodef{STD}[STD]{Short-Term Depression}
\acrodef{STP}[STP]{Short-Term Plasticity}
\acrodef{STT-MRAM}[STT-MRAM]{Spin-Transfer Torque Magnetic Random Access Memory}
\acrodef{STT}[STT]{Spin-Transfer Torque}
\acrodef{SNR}[SNR]{Signal-to-noise ratio}
\acrodef{SW}[SW]{Software}
\acrodef{TCAM}[TCAM]{Ternary Content-Addressable Memory}
\acrodef{TFT}[TFT]{Thin Film Transistor}
\acrodef{USB}[USB]{Universal Serial Bus}
\acrodef{VHDL}[VHDL]{VHSIC Hardware Description Language}
\acrodef{VLSI}[VLSI]{Very Large Scale Integration}
\acrodef{VOR}[VOR]{Vestibulo-Ocular Reflex}
\acrodef{WTA}[WTA]{Winner-Take-All}
\acrodef{WCST}[WCST]{Wisconsin Card Sorting Test}
\acrodef{XML}[XML]{eXtensible Mark-up Language}
\acrodef{divmod3}[DIVMOD3]{divisibility of a number by three}
\acrodef{hWTA}[hWTA]{Hard Winner-Take-All}
\acrodef{sWTA}[sWTA]{soft Winner-Take-All}
\acrodef{DYNAP}[DYNAP]{Dynamic Neuromorphic Asychronous Processor}

\begin{document}

\title{An ultra-low-power sigma-delta neuron circuit\thanks{This work was supported by SNSF grant number $CRSII2\_160756$. \textcopyright 2019 IEEE.  Personal use of this material is permitted.  Permission from IEEE must be obtained for all other uses, in any current or future media, including reprinting/republishing this material for advertising or promotional purposes, creating new collective works, for resale or redistribution to servers or lists, or reuse of any copyrighted component of this work in other works.}}

\author{\IEEEauthorblockN{Manu V Nair and Giacomo Indiveri}	\\ \IEEEauthorblockA{Institute of Neuroinformatics,
		University of Zurich and ETH Zurich\\
		Email: [mnair$|$giacomo]@ini.uzh.ch}}
\maketitle

\begin{abstract}
  Neural processing systems typically represent data using \ac{LIF} neuron models that generate spikes or pulse trains at a rate proportional to their input amplitudes. This mechanism requires high firing rates when encoding time-varying signals, leading to increased power consumption. Neuromorphic systems that use adaptive \ac{LIF} neuron models overcome this problem by encoding signals in the relative timing of their output spikes rather than their rate. In this paper, we analyze recent adaptive \ac{LIF} neuron circuit implementations and highlight the analogies and differences between them and a first-order \sd feedback loop. We propose a new \sd neuron circuit that addresses some of the limitations in existing implementations and present simulation results that quantify the improvements. We show that the new circuit, implemented in a $1.8 V$, $180 nm$ CMOS process, offers up to $42 dB$ \ac{SDR} and consumes orders of magnitude lower energy. Finally, we also demonstrate how the sigma-delta interpretation enables mapping of real-valued \acp{RNN} to the spiking framework to emphasize the envisioned application of the proposed circuit.
\end{abstract}

\begin{IEEEkeywords}
\sd, recurrent neural networks, circuit, neuromorphic
\end{IEEEkeywords}

\section{Introduction}
\label{sec:introduction}
The effectiveness of artificial neural networks in pattern recognition and classification tasks makes it compelling to use them in ultra-low power applications such as biomedical implants or energy-harvesting smart sensors. However, the computation of neural network models on traditional \vn style processors tends to consume considerable amounts of energy, making them nonviable in such power-starved conditions. Event-based neuromorphic processing is an alternative computational approach that tries to address this problem by a combination of in-memory and asynchronous communication techniques \cite{Aamir_etal18brainscales, Chicca_etal14, Benjamin_etal14, Payvand_etal18faraday}.

Event-based neuromorphic systems process signals by encoding them as a sequence of asynchronous spikes. The most commonly-used encoding mechanism is rate-coding, where the firing-rate of a \ac{LIF} neuron is proportional to the  amplitude of the input signal. In these conditions, decoding a signal is achieved by simply low-pass filtering the corresponding spike train. To achieve good transmission quality with temporally-changing inputs, it is necessary to set the \ac{LIF} neuron's firing rate sufficiently high to encode the input within a short time window. However, transporting a large number of spikes consumes large amounts of power.

Biology has found a solution to this problem by endowing neurons with \emph{spike frequency adaptation} mechanisms~\cite{Connors_etal82, Brette_Gerstner05}. A neuron model that implements such a mechanism is the \ac{ADEX} one. This model has been implemented in silicon in several neuromorphic processors ~\cite{Millner_etal10,Livi_Indiveri09, Indiveri03a, Chicca_etal14,Qiao_etal15,Indiveri_etal11}. In this paper, we observe that the \ac{ADEX} neuron model can be interpreted as a first-order \sd loop. We analyze the \ac{ADEX} silicon neuron circuits in this light and identify issues that affect their power consumption and encoding quality. We then describe an improved \sd neuron circuit that addresses these issues,  dramatically lowering power consumption and improving the signal encoding quality. Finally, we present a recurrent neural network simulation to highlight that the \sd interpretation allows us to map floating-point implementation of a recurrent neural network to a spiking neuromorphic one.

\section{Neuromorphic signal chain}
\label{sec:neur-sign-chain}

\begin{figure}[!ht]
	\centering
	\includegraphics[width=\linewidth]{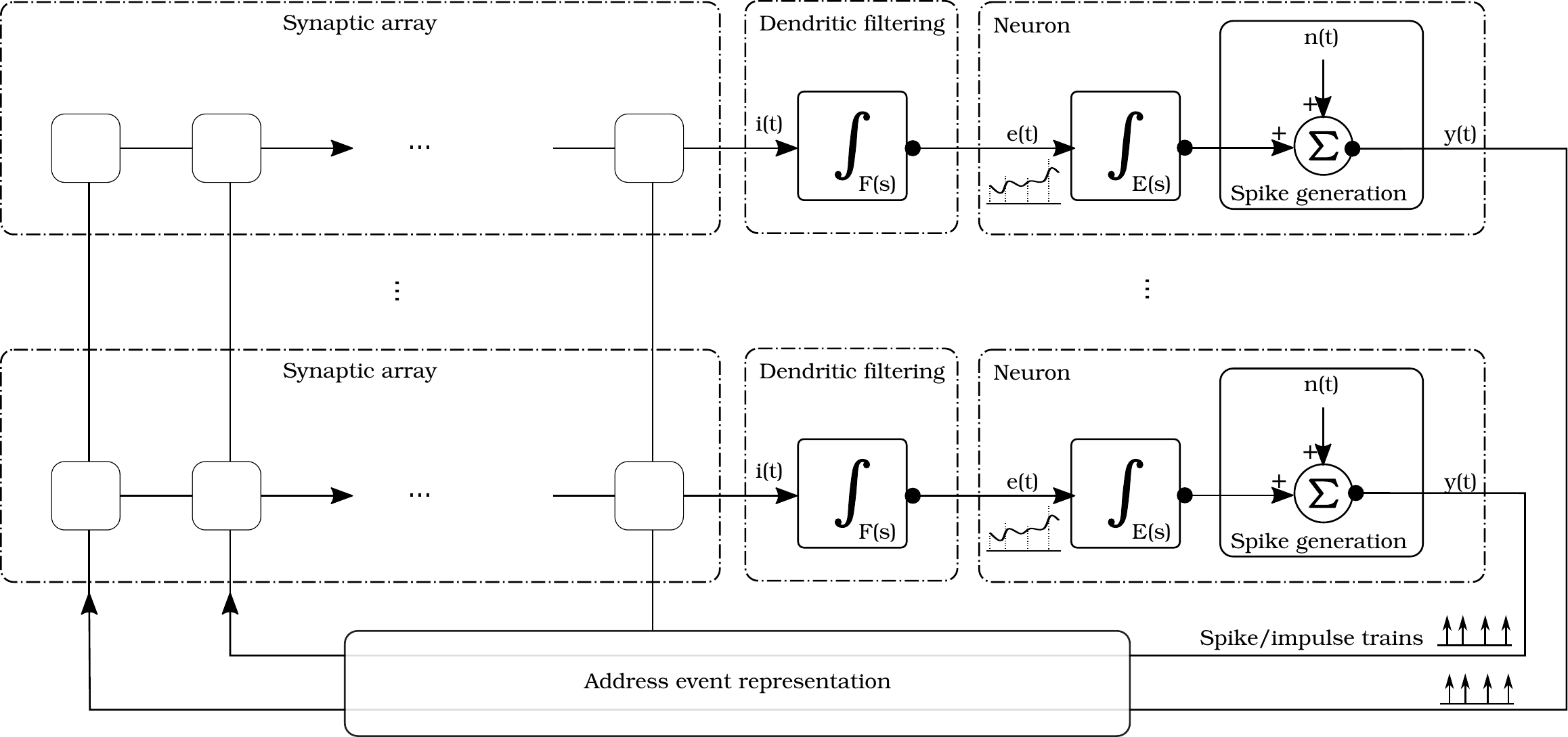}
	\caption{ Signal chain in a spiking neuromorphic system.}
	\label{arch}
\end{figure}

Fig.~\ref{arch} illustrates the signal flow in a typical recurrently connected neuromorphic system. In this scheme, spike trains generated by the neurons are weighted by the synapse circuits. The weighted sum of currents, $i(t)$ is then integrated by the dendrite circuit F(s). The net synaptic current is further filtered by a neuron leak integrator circuit E(s) before reaching the neuron spike generation block that generates a voltage pulse when its input exceeds a predefined threshold value. The spike generation mechanism introduces non-linearities, $n(t)$ in the signal encoded as a spike train that can be reduced by increasing the firing rate of the neurons.
%These distortions are analogous to the quantization noise introduced in \ac{ADC} circuits. Increasing the firing rate is therefore analogous to increasing the sampling rate.

\section{The \ac{ADEX} model as a \sd encoder}
\label{sec:acadex-circuit}
\ac{ADEX} neuron circuits, such as the one presented in~\cite{Chicca_etal14} typically use a feedback mechanism to model the spike frequency adaptation mechanism observed in real neurons, as described by the following equations:
\begin{align}
\tau_{mem} \frac{dI_{mem}}{dt} =& -\alpha_L \cdot (I_{mem} - I_L) \label{adex1} \\
\nonumber &+ \alpha_L \cdot \Delta_T \cdot exp(\frac{I_{mem}-\delta}{\Delta_T}) - s + i \\
\tau_w \frac{ds}{dt} =& \alpha_s (I_{mem}-I_L) - s
\label{adex}
\end{align}
where, $I_{mem}$ and $I_L$  are the ``membrane potential'' and ``leak reversal potential'' variables that are represented as currents. The term $s$ represents the adaptation current, $i$ the input current, $\tau_{mem}$ the membrane time constant, $\alpha_L$ a gain factor, $\delta$ the threshold, $\Delta_T$ the slope factor, $\alpha_s$ the adaptation coupling parameter and $\tau_w$ is the adaptation time constant. The variable $I_{mem}$ is reset to its resting value when $I_{mem} > \delta$.
\begin{figure}[!ht]
	\centering
	\includegraphics[width=0.8\linewidth]{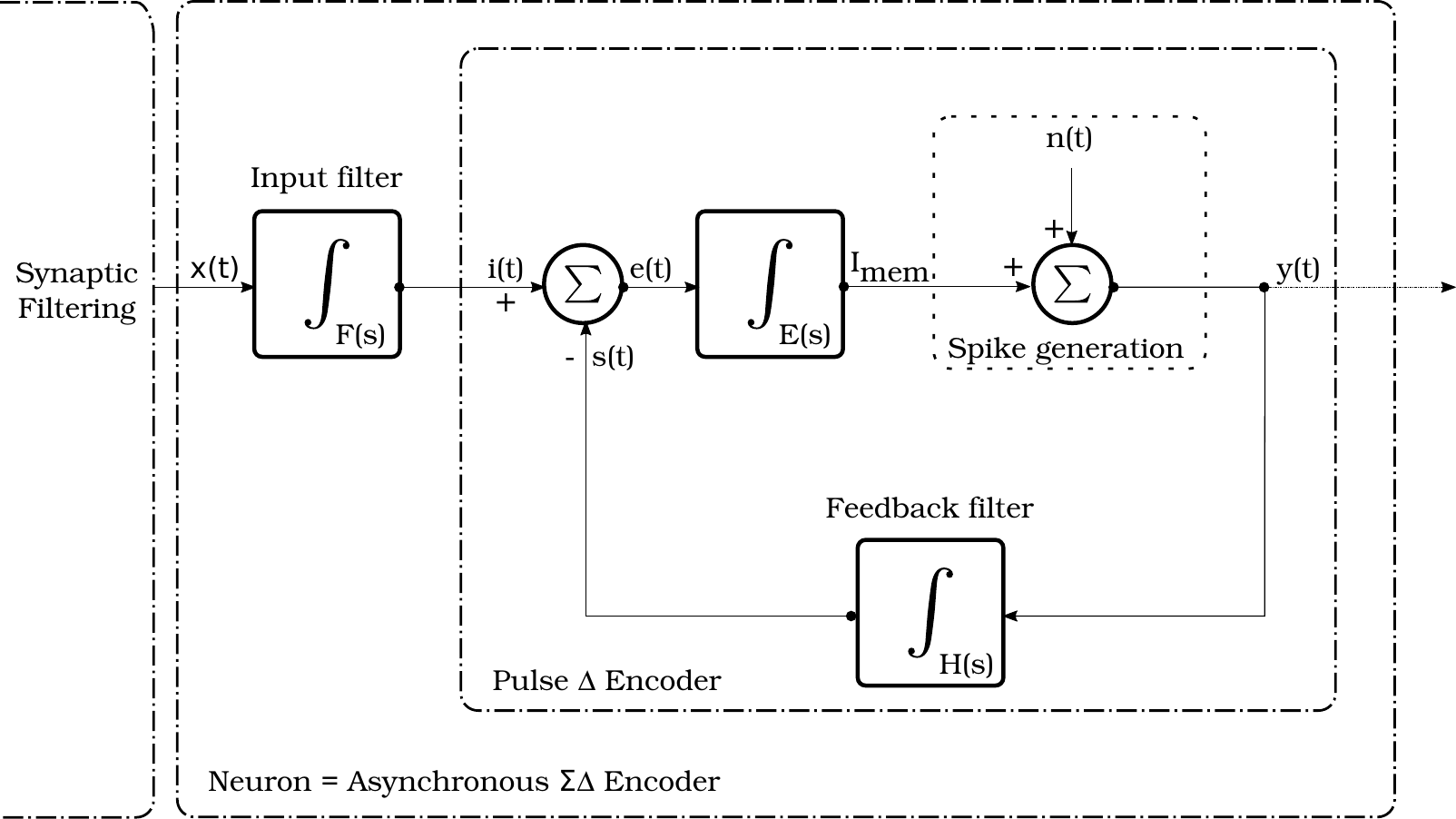}
    \caption{ Block diagram of a \sd neuron circuit. }
    \label{fig:sd_blockdiag}
\end{figure}
As $\alpha_s$ in the \ac{ADEX} model is set to a very small value, the changes in $s(t)$ are dominated by the spike events: the change in $s(t)$ is that of a first-order low-pass filter responding to a train of impulses. This filter is marked $H(s)$ in Fig.~\ref{fig:sd_blockdiag}.
The difference between the input current $i(t)$ and the feedback signal $s(t)$ is filtered by a first-order filter $E(s)$ with gain, $\alpha_L$, and time constant, $\tau_{mem}$. When the output of this filter, $I_{mem}$, exceeds the neuron spiking threshold, a spike is generated. This is modeled by the exponential term in Equation~\ref{adex1}.
With every spike $s(t)$ increases and the  difference between the $i(t)$ and $s(t)$ decreases, thus completing the feedback loop. This mechanism is equivalent to an atypical continuous-time $\Delta$ modulation loop \cite{Pavan_etal17} with the difference being that the output spikes are unipolar. As described in Section~\ref{sec:neur-sign-chain}, input signals arriving to the neuron are encoded as spikes trains and  pre-filtered by dendritic or synaptic stages. This filtering operation, $F(s)$ can also be interpreted as a $\Sigma$ stage and the combination of the $\Sigma$ stage and $\Delta$ modulator can be interpreted as a first-order  \sd loop. A similar interpretation of the spike response model\cite{Gerstner_Kistler02} in neuroscience has also been proposed~\cite{Yoon17}.

Note that the neuron model only generates a spike when the filtered difference or error reaches a spiking threshold. The generated spike trains can be low-pass filtered (for instance, by $F(s)$) to reconstruct the output in an asynchronous manner. There is no quantization in any stage of the signal chain and therefore, no quantization noise is introduced into the signal~\cite{Ouzounov_etal06}. However, aliasing and non-linear effects~\cite{Ouzounov_etal06, Roza97} do affect the encoding quality. Depending on the first-order leak to avoid using bipolar pulses also introduces other issues into the loop that are not seen in typical \sd loop. For example, the limit cycle frequency of the proposed loop is proportional to the amplitude of the input signal. In the absence of an input, there is no spike. The gain and time constant of the feedback filters also affect inputs of different frequencies and amplitudes differently. %Some of these issues will be highlighted in the simulation section.

\section{The \ac{ADEX} and \sd neuron circuits}
\label{sec:rolls}
The circuit shown in Fig.~\ref{fig:rolls} is a neuromorphic implementation of the $\Delta$ modulator block of Fig.~\ref{fig:sd_blockdiag} that is adapted from commonly used silicon neuron circuits in the literature~\cite{Chicca_etal14,Aamir_etal18brainscales}. It is a current-mode circuit where the filter $E(s)$ integrates the difference between the input and feedback outputs. When this difference exceeds the switching threshold of the inverter, it generates a spike. This spike-event induces an increase in the charge held by the feedback capacitance, $C_{fb}$, which in turn increases the feedback current closing the feedback loop. Note that the filters $E(s)$ and $H(s)$ are implemented using \ac{DPI} circuits \cite{Chicca_etal14}.

\begin{figure}[!ht]
	\centering
	\includegraphics[width=0.8\linewidth]{rolls}
	\caption{ Circuit schematics of the \ac{ADEX} neuron.}
	\label{fig:rolls}
\end{figure}

%The most serious problem with the circuit is the input to the feedback filter.
For accurate reconstruction in the \sd loop, it is essential that the pulse trains used for reconstructing the signals be the same as that used in the feedback filter. However, the spike integrated by the feedback filter in the \ac{ADEX} neuron circuit is a narrow digital pulse, which does typically not last long enough for the \ac{DPI} feedback filter to produce a sufficiently large output. The sub-threshold \ac{DPI} feedback filter produces a weak response to this narrow pulse. To recover from this, the neuron is forced to spike much more frequently causing the circuit to slew, decreasing performance and increasing power consumption. In this state, the circuit behaves more like a \ac{LIF} neuron than a \sd loop.

\begin{figure}[!ht]
	\centering
	\includegraphics[width=0.8\linewidth]{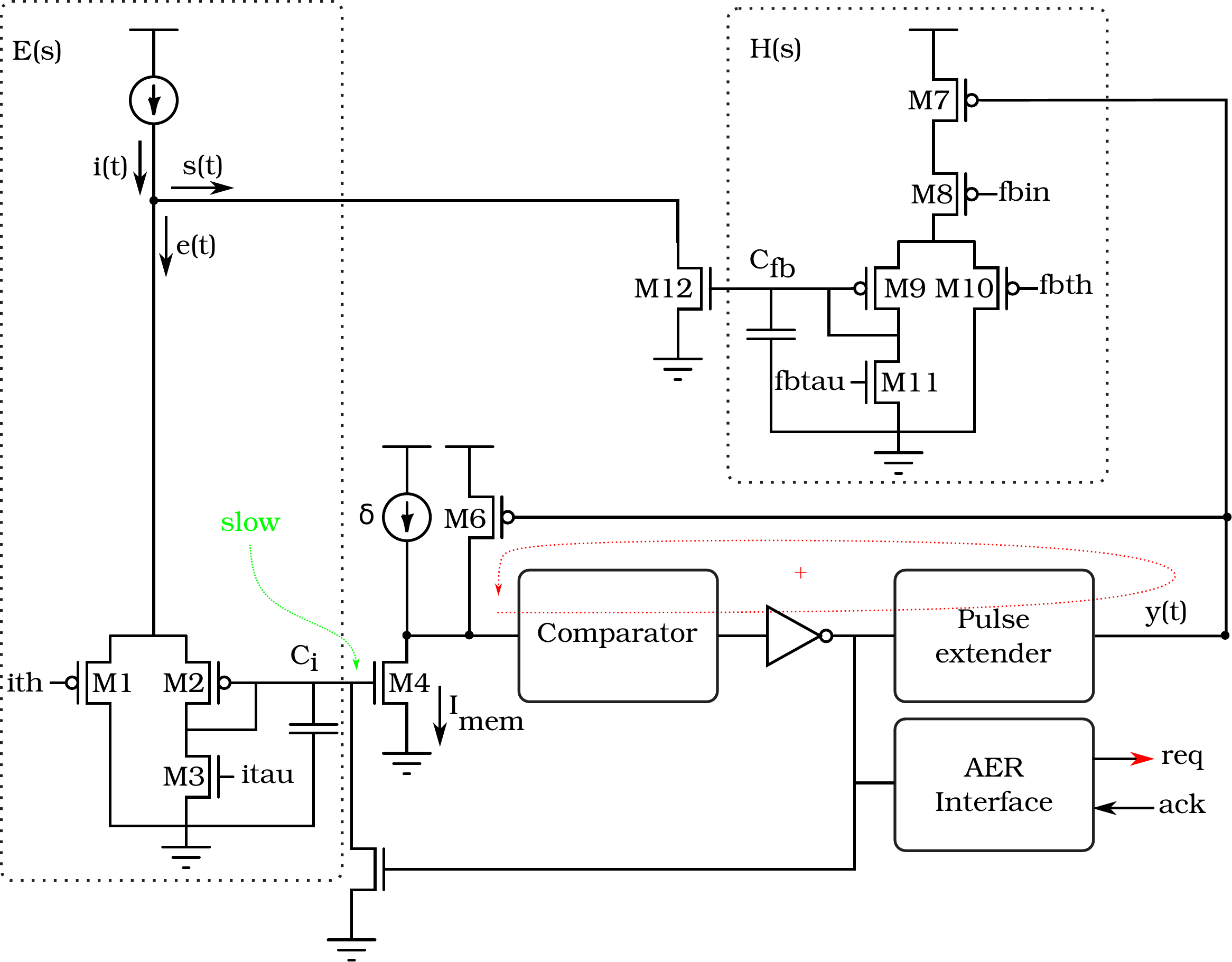}
	\caption{ Circuit schematics of the \sd neuron.}
	\label{fig:sd_block}
\end{figure}

The second problem that mainly affects the energy consumption of the neuron is that the comparator circuit used for generating spikes is an inverter, with a switching threshold of approximately $0.5V_{DD}$. $I_{mem}$ is a slow-changing signal and as long as $I_{mem}$ is slowly rising, the inverter is constantly sinking current. Moreover, the large switching threshold drives the $E(s)$ \ac{DPI} circuit out of sub-threshold causing non-linear distortions. To address these issues, we modify the circuit by adding a pulse-extender to the spike-generator block and introducing a new low-power current comparator to determine when the integrated current exceeds the (now tunable) spiking threshold. The new \sd circuit schematic is  shown in Fig.~\ref{fig:sd_block}.

\subsection{Starved current comparator circuit}
\begin{figure}[!ht]
	\centering
	\subfloat[Current comparator \label{fig:cc}]{\includegraphics[width=0.4\linewidth]{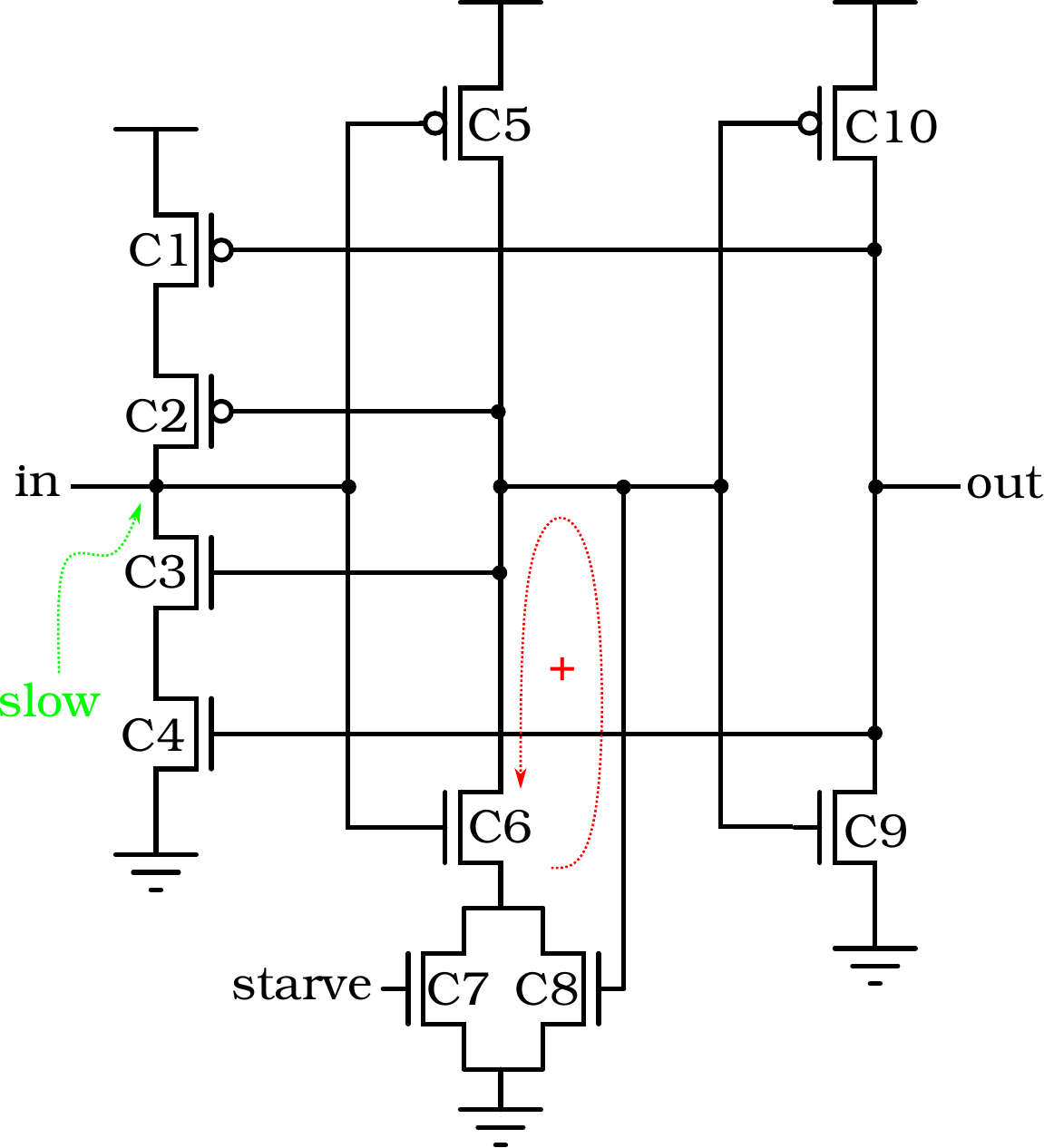}}
  \subfloat[Pulse extender\label{fig:pe}]{\includegraphics[width=0.6\linewidth]{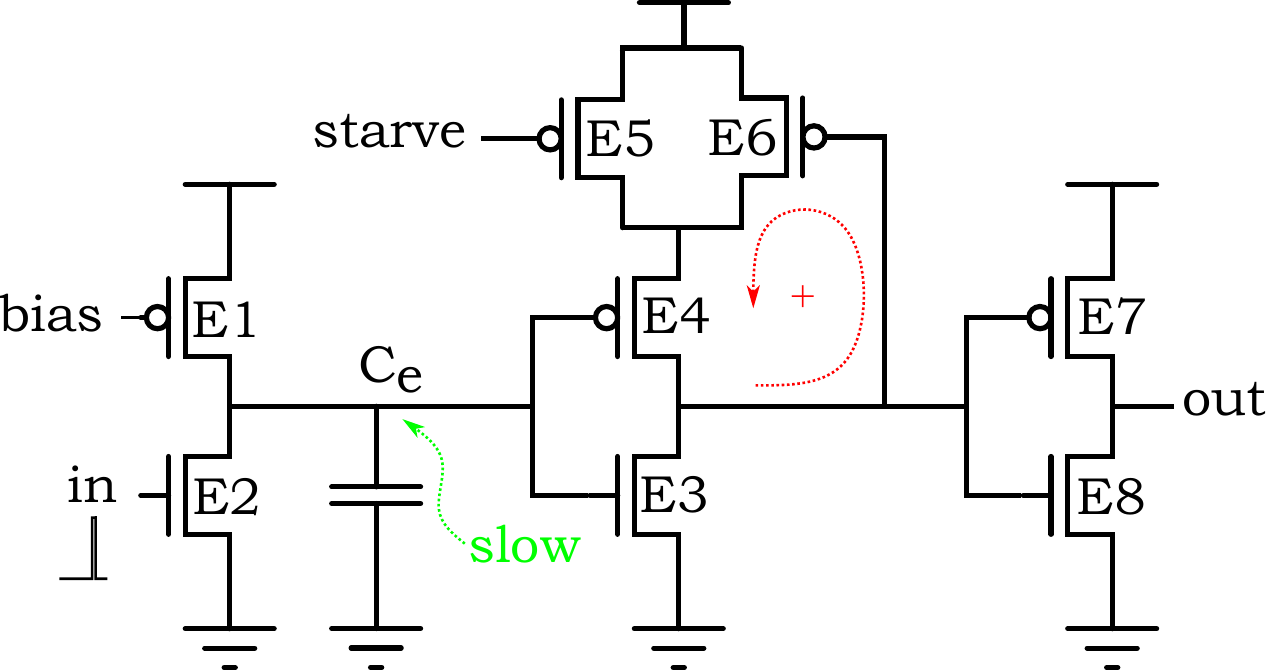}} \\
	\caption{Circuit schematics of the improved \sd neuron blocks}
	\label{fig:ckts}
\end{figure}

The comparator circuit (Fig.~\ref{fig:cc}) is a current-mode circuit based on \cite{Banks_Toumazou08}, with two additional transistors $C7$ and $C8$ added to save power. The input to the comparator circuit is set by the competition between the currents representing the spiking threshold, $\delta$, and $I_{mem}$. When $I_{mem}$ is smaller than $\delta$, the input node to the comparator is close to $V_{DD}$ and vice versa. $I_{mem}$ is a slow-changing signal and so the voltage at ``in'' falls slowly as the $I_{mem}$ approaches the  $\delta$. During this time $C7$ only permits a small current set by the $starve$ bias and $C8$ is off reducing the power wasted through the $C5-C6$ inverter. When the input drops sufficiently, $C8$ turns on and initiates fast switching. This spike event is fed to the pulse extender. The starved current-comparator circuit is responsible for the dramatic reduction in power consumed by the new neuron circuit. A bonus feature is that it allows the spiking threshold of the neuron to be tuned.

\subsection{Pulse extender}
The $C_e$ node of the pulse extender (Fig.~\ref{fig:pe}) normally sits at $V_{DD}$. As soon as the comparator circuit generates a spike event, the capacitor node is discharged. Immediately after the capacitor is discharged, the comparator input is reset. The time between the generation of a spike event from the comparator to resetting its input is very short. The role of the pulse extender is to stretch the spike event for use in the feedback filter. The circuit achieves this by slowly charged $C_e$ by a current source $E1$. When it rises sufficiently, the output switches back to its resting condition. The same bias current is shared by all the pulse extender blocks in the neuromorphic array ensuring that all the filters in the system operate on similar pulses improving the transmission accuracy.

\subsection{Regenerative feedback saves power}
As soon as the pulse extender registers a spike event, the capacitor $C_i$ is fully discharged. This saves power by preventing wasteful current discharge through the $C5-C6$ inverter stage. The pulse extender output also pulls up the input node of the comparator using a regenerative feedback mechanism through $M6$ in Fig.~\ref{fig:sd_block}. In the absence of this regenerative feedback loop, the comparator input node would be charged only by a small spiking threshold current increasing the power consumption of comparator transistor pair $C5-C6$ very significantly. Similar regenerative feedback loops are also used in the comparator and pulse extender circuits as highlighted with red loops in the circuit diagrams. Any circuit, neuromorphic or otherwise, that interfaces a comparator-like circuit to a slow-changing input could benefit from this technique.

%\begin{figure}[!ht]
%	\centering
%	\includegraphics[width=\linewidth]{dpi_snr}
%	\caption{Signal to distortion ratio of the DPI block used to in the neuron circuit for different values of gain and input frequency. The \ac{DPI} circuit was setup with a pole at 10 Hz..}
%	\label{fig:dpi_snr}
%\end{figure}

\section{Simulations}
\subsection{Regenerative feedback }
\begin{figure}[!ht]
	\centering
	\subfloat[Maximum firing rate\label{fig:max_fire}]{\includegraphics[width=0.5\linewidth]{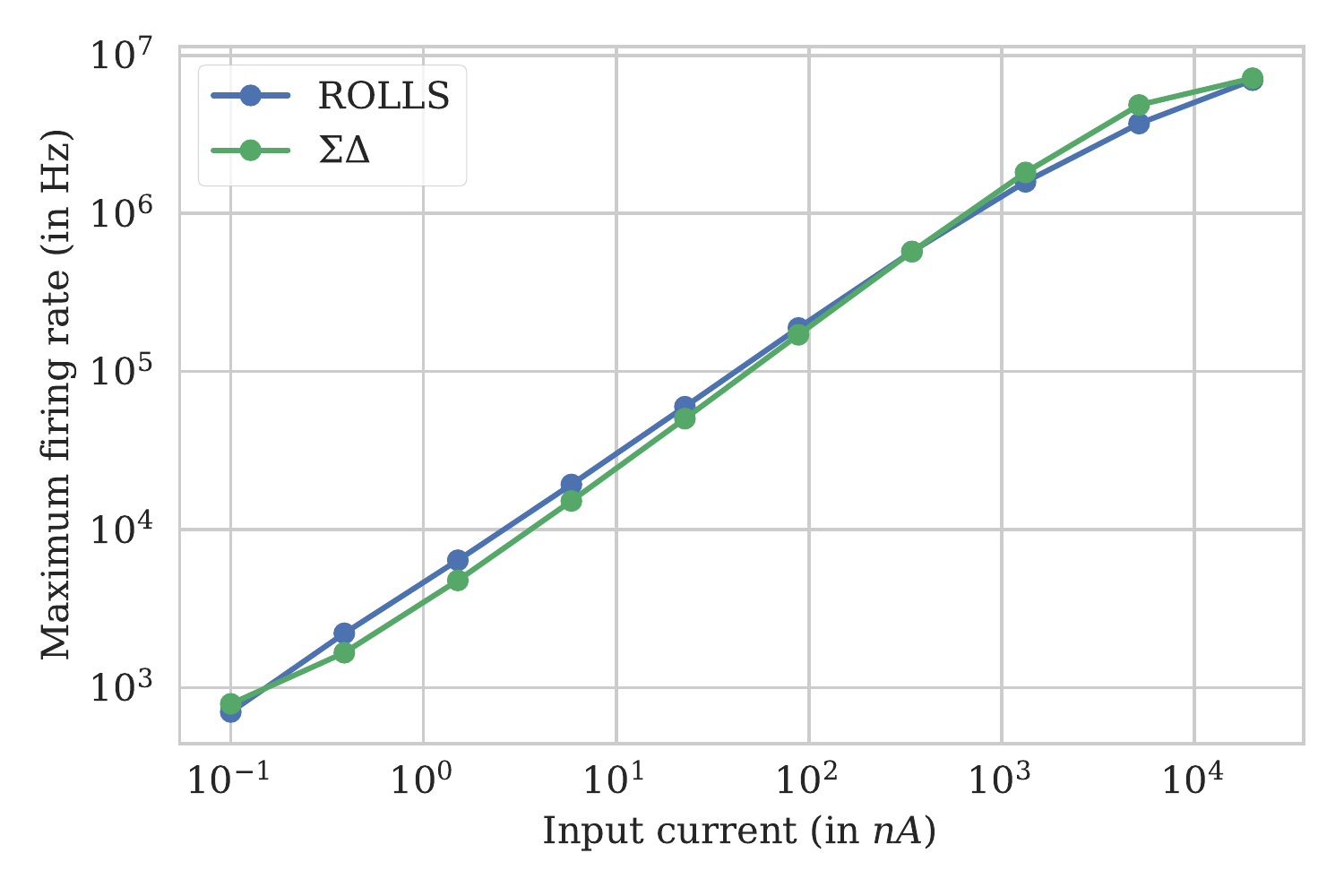}}
	\subfloat[Energy consumed per spike\label{fig:max_fire_energy}]{\includegraphics[width=0.5\linewidth]{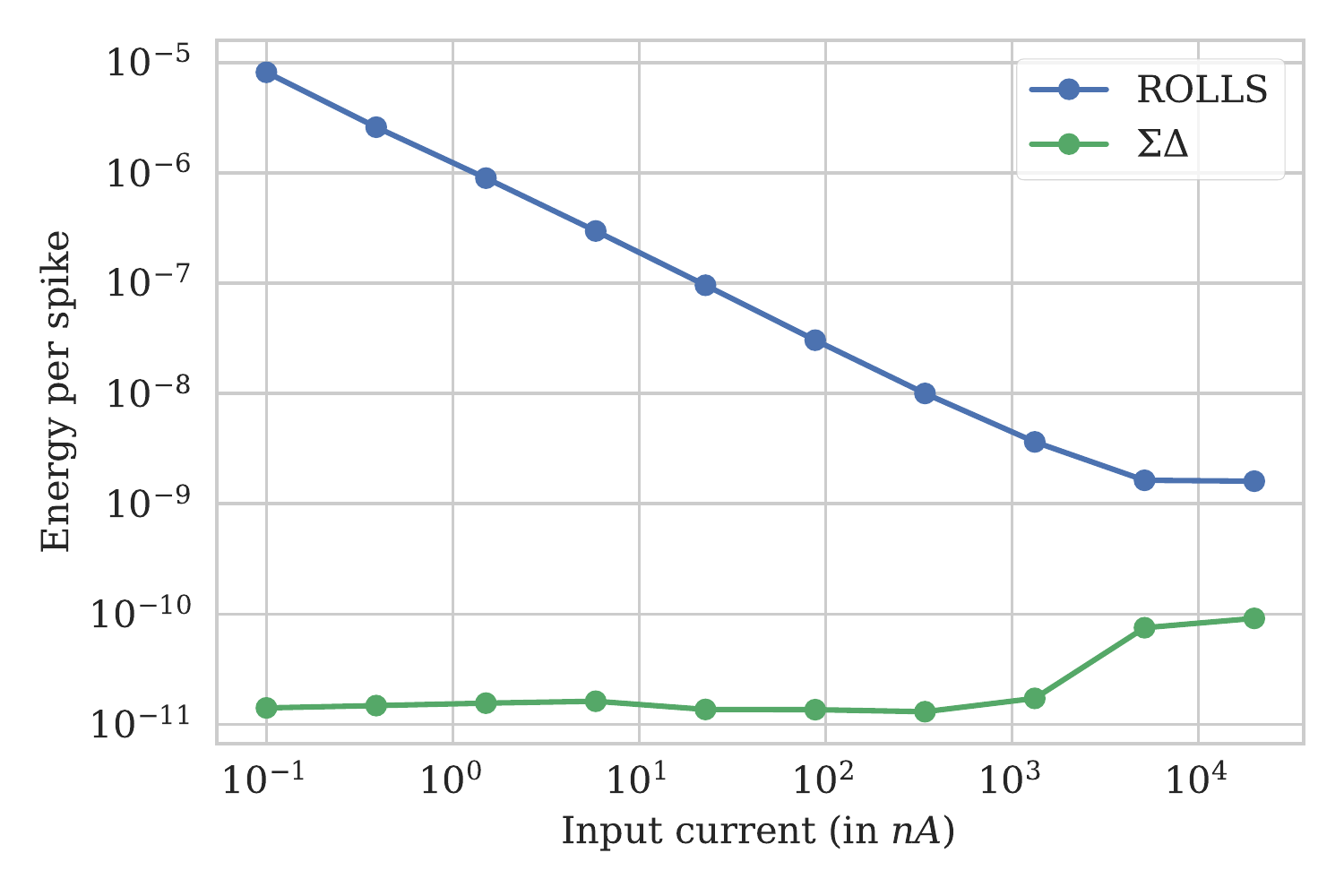}}\\
	\subfloat[\ac{ADEX} circuit\label{fig:rolls_energy}]{\includegraphics[width=0.5\linewidth]{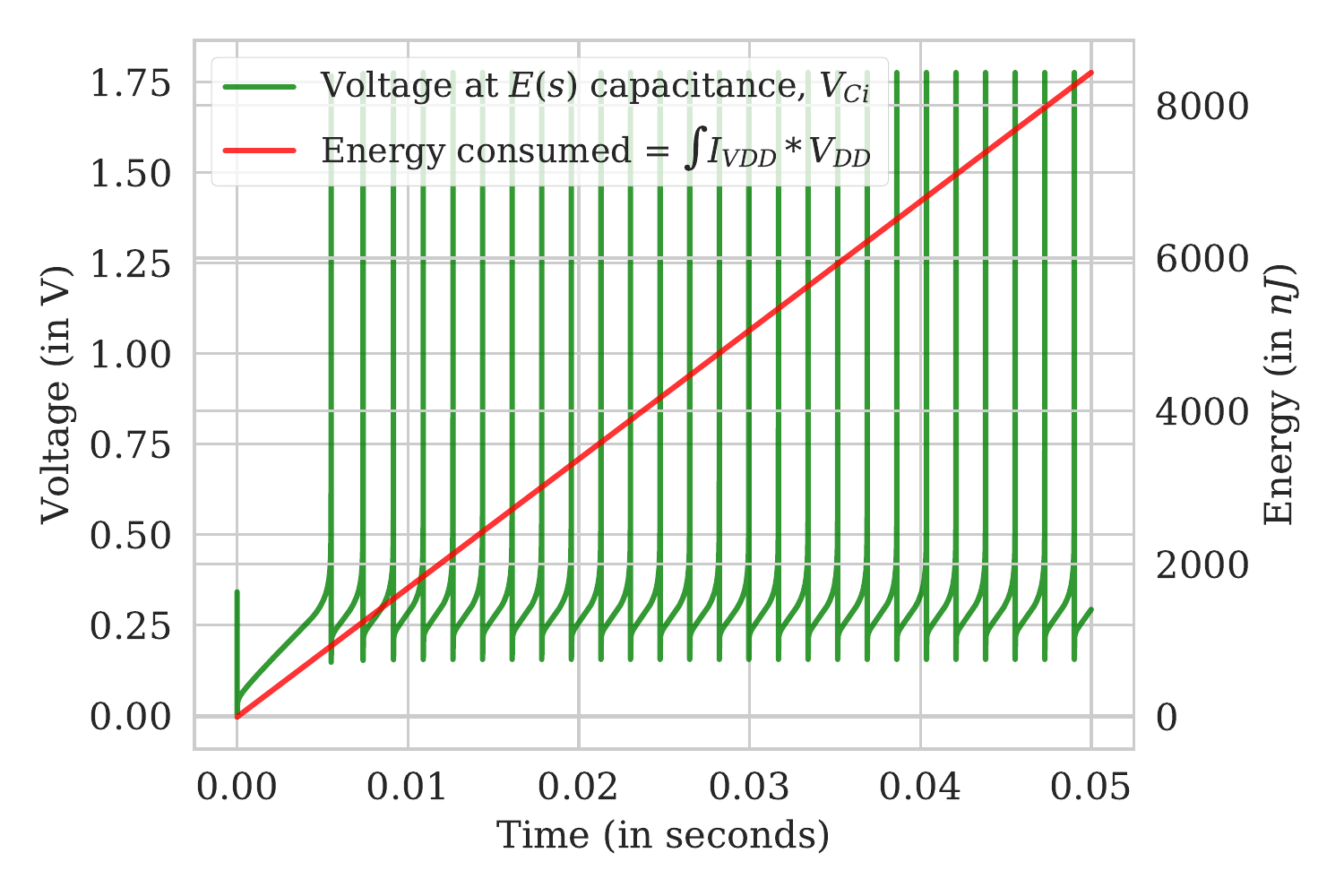}}
    \subfloat[\sd neuron circuit \label{fig:sd_energy}]{\includegraphics[width=0.5\linewidth]{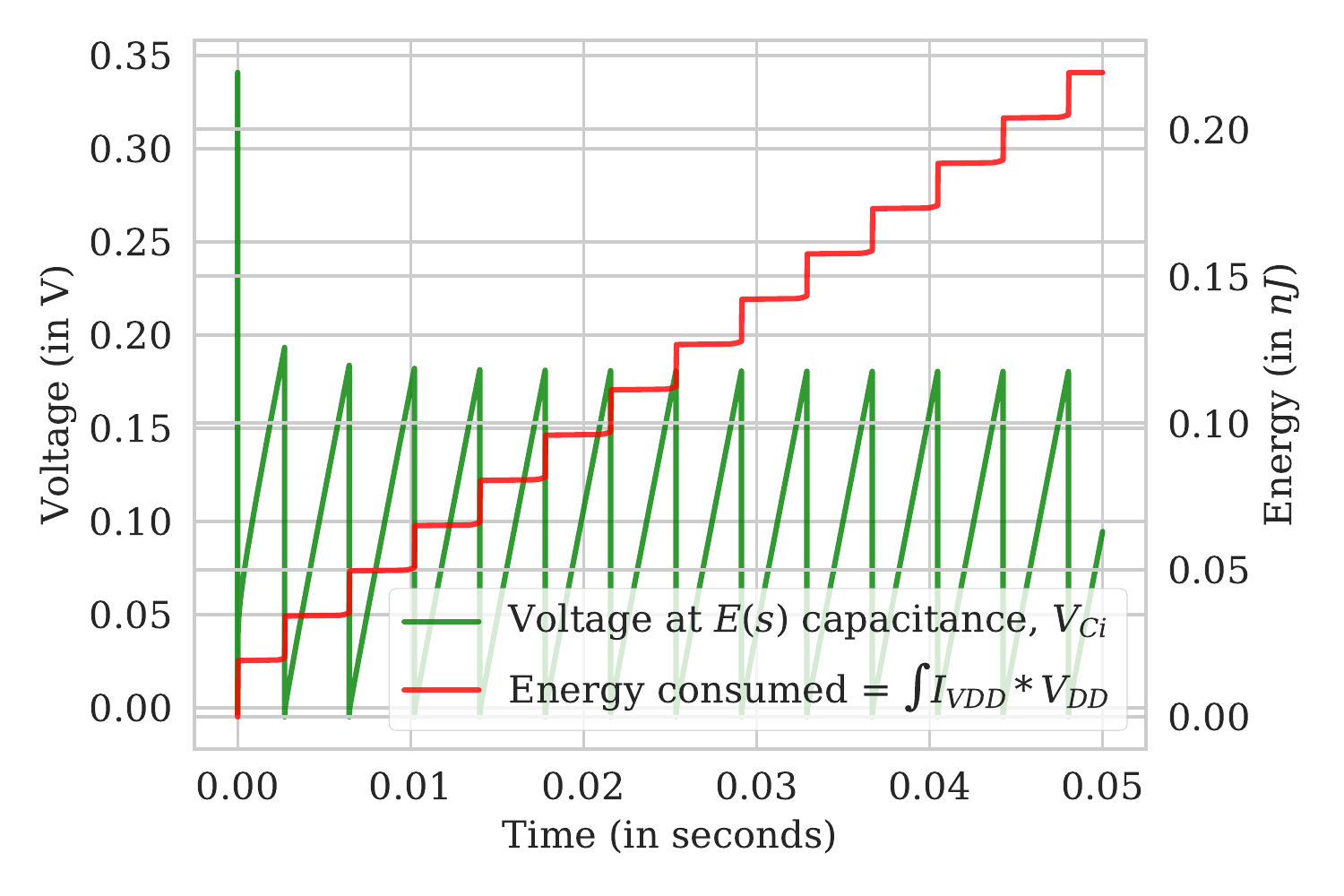}}\\
	\caption{Simulations for DC input. For \ref{fig:rolls_energy} and \ref{fig:sd_energy}, a $50 pA$ DC input was applied. $ \text{Energy per spike} = \frac{\text{Total energy consumed}}{\text{firing rate} * \text{simulation time}} $.}
	\label{fig:max_firing}
\end{figure}

To isolate the improved energy-efficiency of the new neuron circuit due to regenerative feedback, we disable the feedback \ac{DPI} block and apply DC input currents. In this mode, both circuits act as \ac{LIF} neurons and we observe a linear relationship between the firing rate and input current~(Fig.~\ref{fig:max_fire}). Observe that the energy consumed per spike in the new \sd circuit is orders of magnitude lower than the \ac{ADEX} model~(Fig.~\ref{fig:max_fire_energy}). This is because of the three regenerative feedback loops in the \sd circuit that are missing in the \ac{ADEX} circuit. The absence of these loops manifests as a continuous drain of energy in the \ac{ADEX} circuit (Fig.~\ref{fig:rolls_energy}). This is in contrast to the step-like increase in energy consumed in the \sd neuron (Fig.~\ref{fig:sd_energy}). In most use-cases, the maximum firing rate of a neuron is limited by the communication bandwidth allocated to the neuromorphic system, typically to values around $10^2 - 10^3 Hz$, where the improved energy-efficiency of the circuit is most useful.

\subsection{Slewing}
\begin{figure}[!ht]
	\centering
	\subfloat[\ac{ADEX} neuron \label{fig:rolls_trans}]{\includegraphics[width=0.5\linewidth]{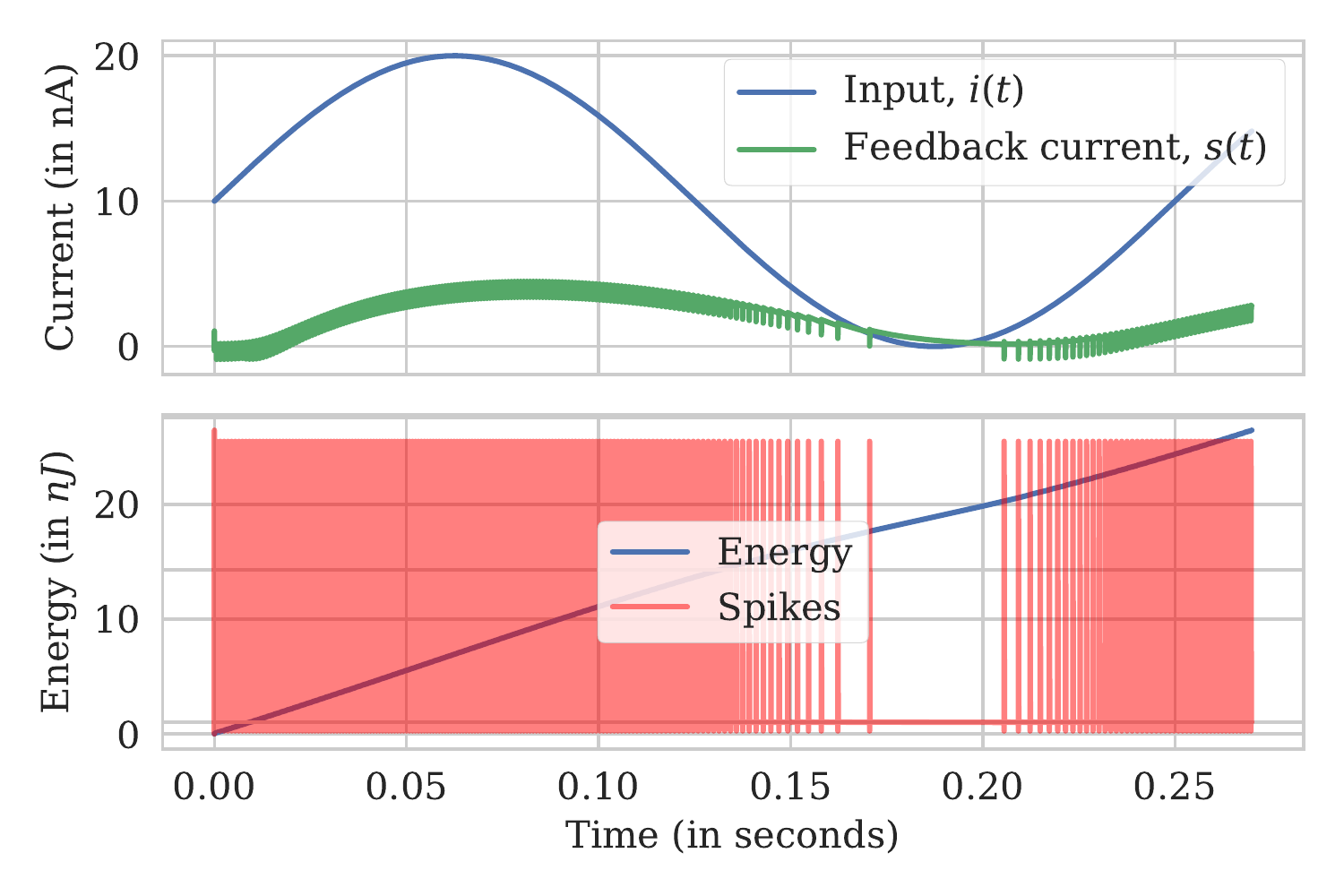}}
	\subfloat[\sd neuron \label{fig:sd_trans}]{\includegraphics[width=0.5\linewidth]{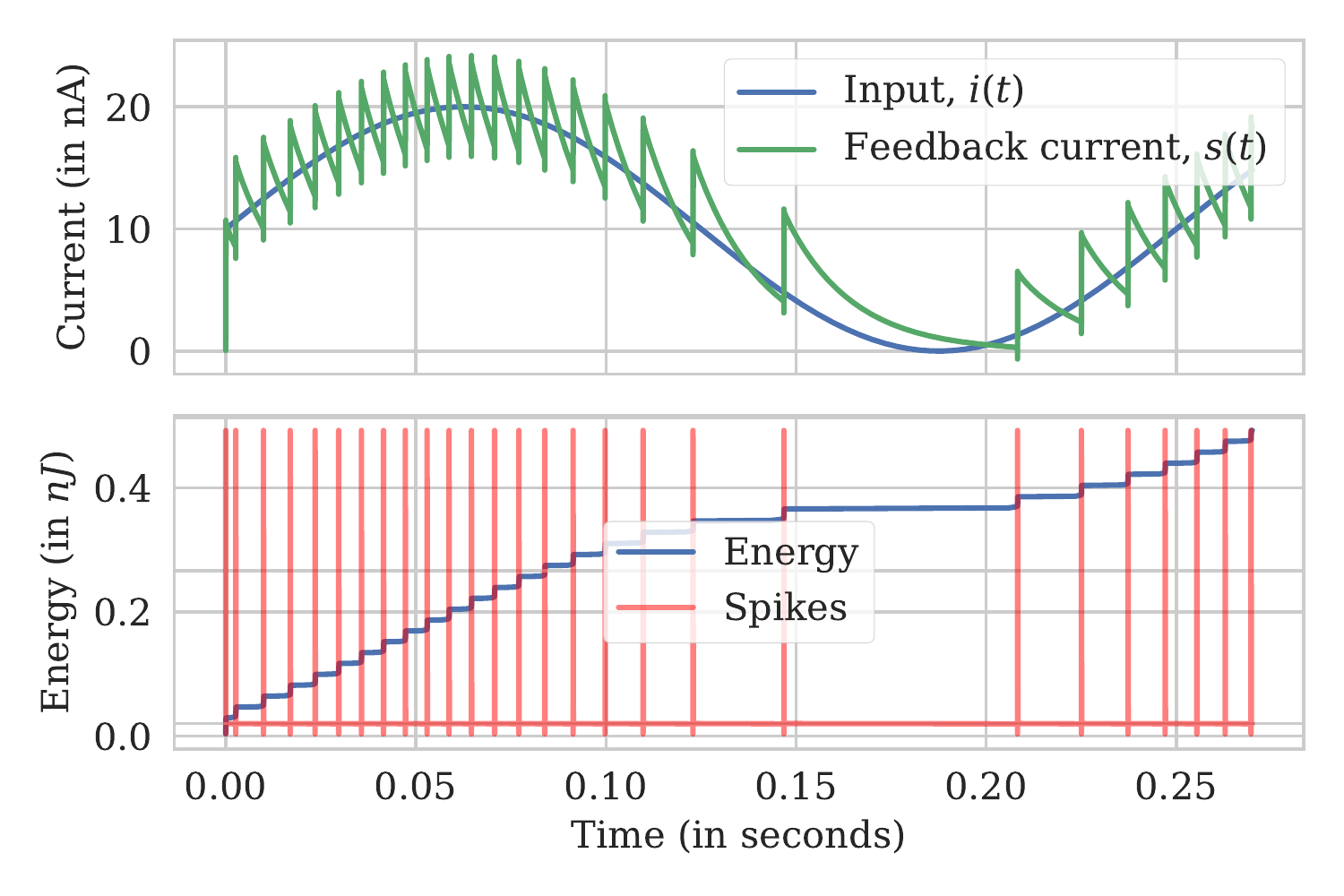}}
	\caption{Sinusoid input response highlights slewing in the \ac{ADEX} neuron circuit.}
	\label{fig:energy_trans}
\end{figure}
Fig.~\ref{fig:energy_trans} highlights slewing in the \ac{ADEX} neuron due to use of short-duration feedback pulses and the resultant increase in firing rate. In contrast, the pulse extender ensures that each spike event is sufficiently long in the \sd circuit.

\subsection{Encoding efficiency}
\begin{figure}[!ht]
	\centering
  \subfloat[\ac{SDR} (\ac{ADEX} neuron) \label{rolls_fbgain_freq_snr}]{\includegraphics[width=0.5\linewidth]{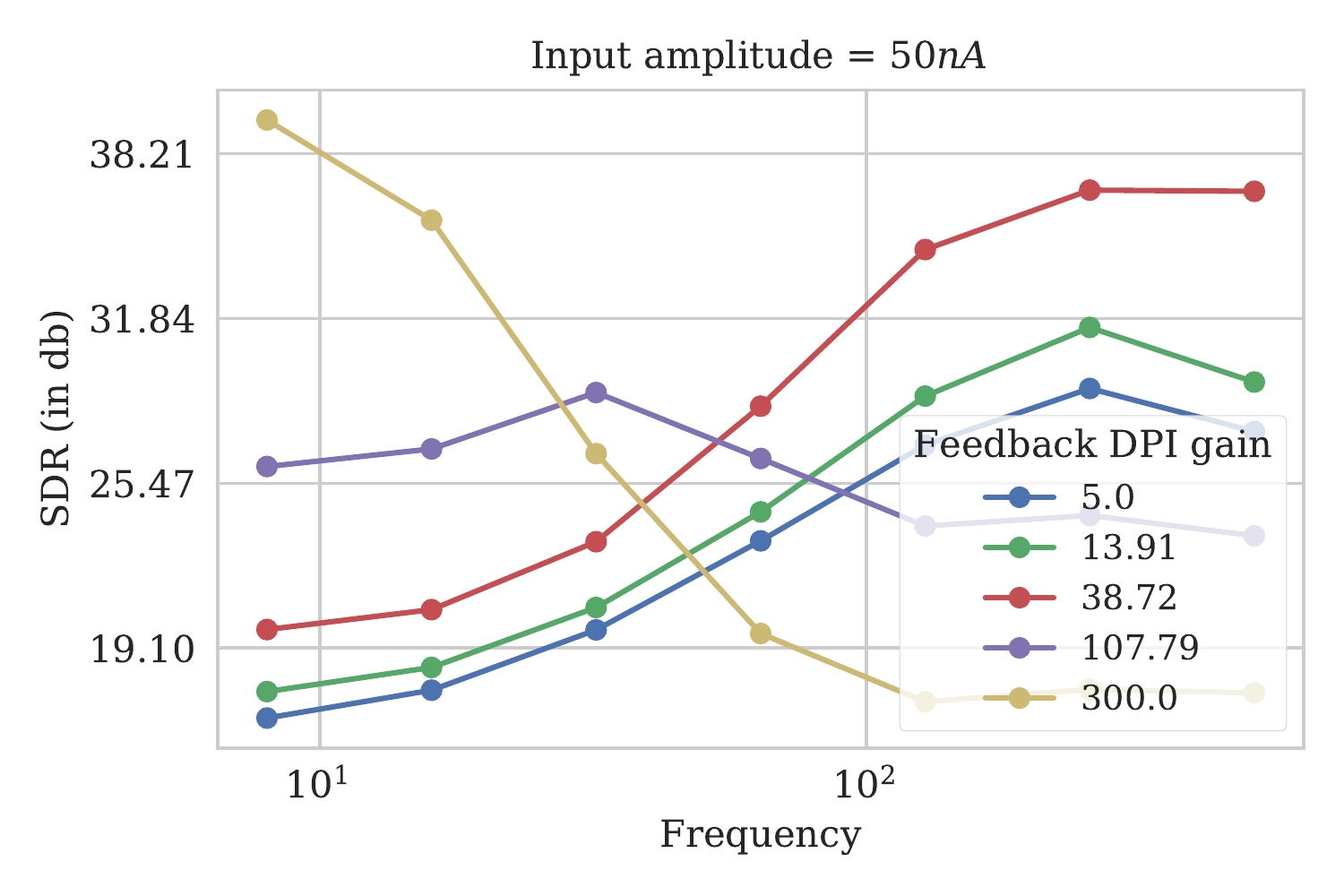}}
  \subfloat[Energy consumed (\ac{ADEX}) \label{rolls_fbgain_freq_power}]{\includegraphics[width=0.5\linewidth]{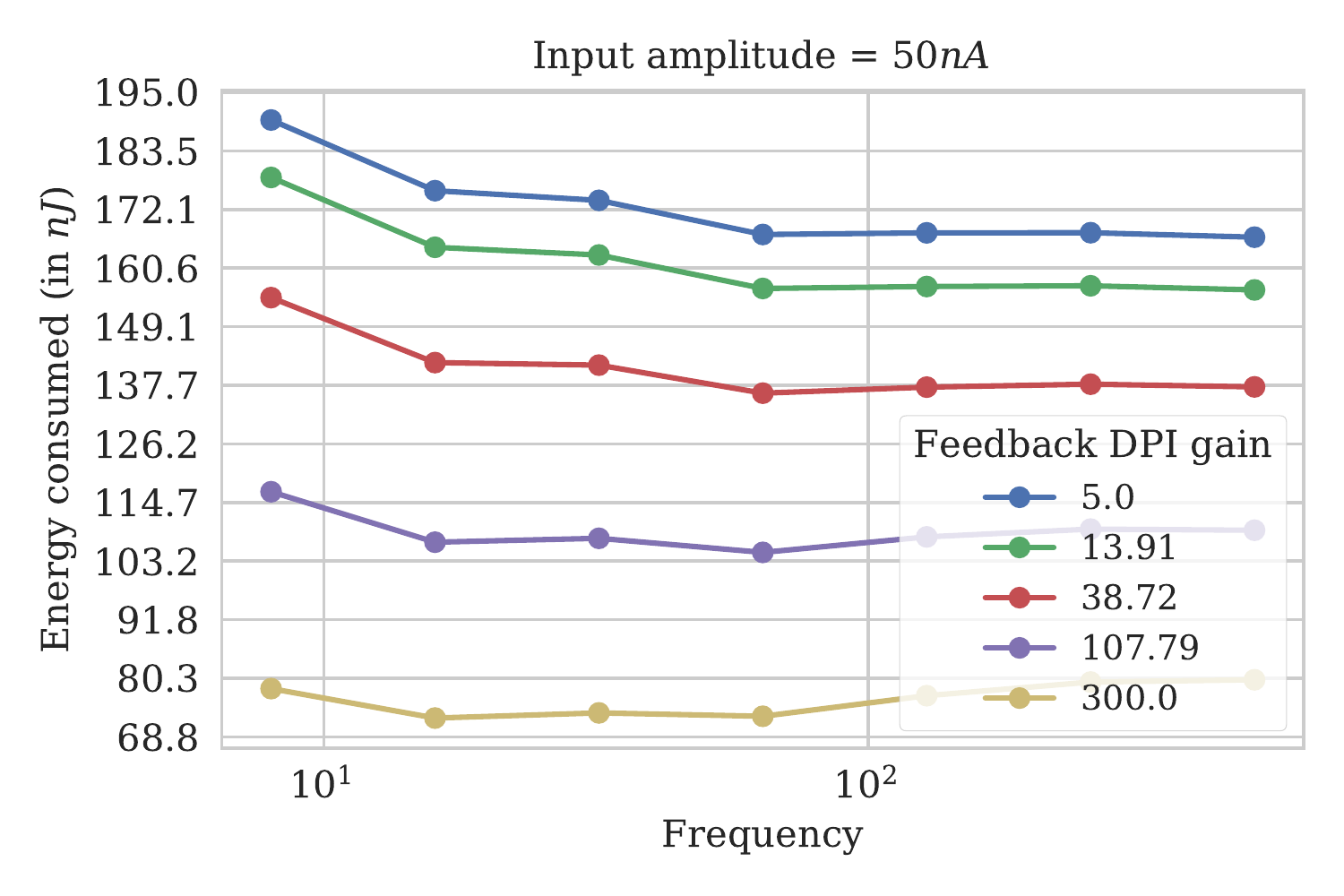}}\\
  \subfloat[\ac{SDR} (\sd neuron) \label{sd_fbgain_freq_snr}]{\includegraphics[width=0.5\linewidth]{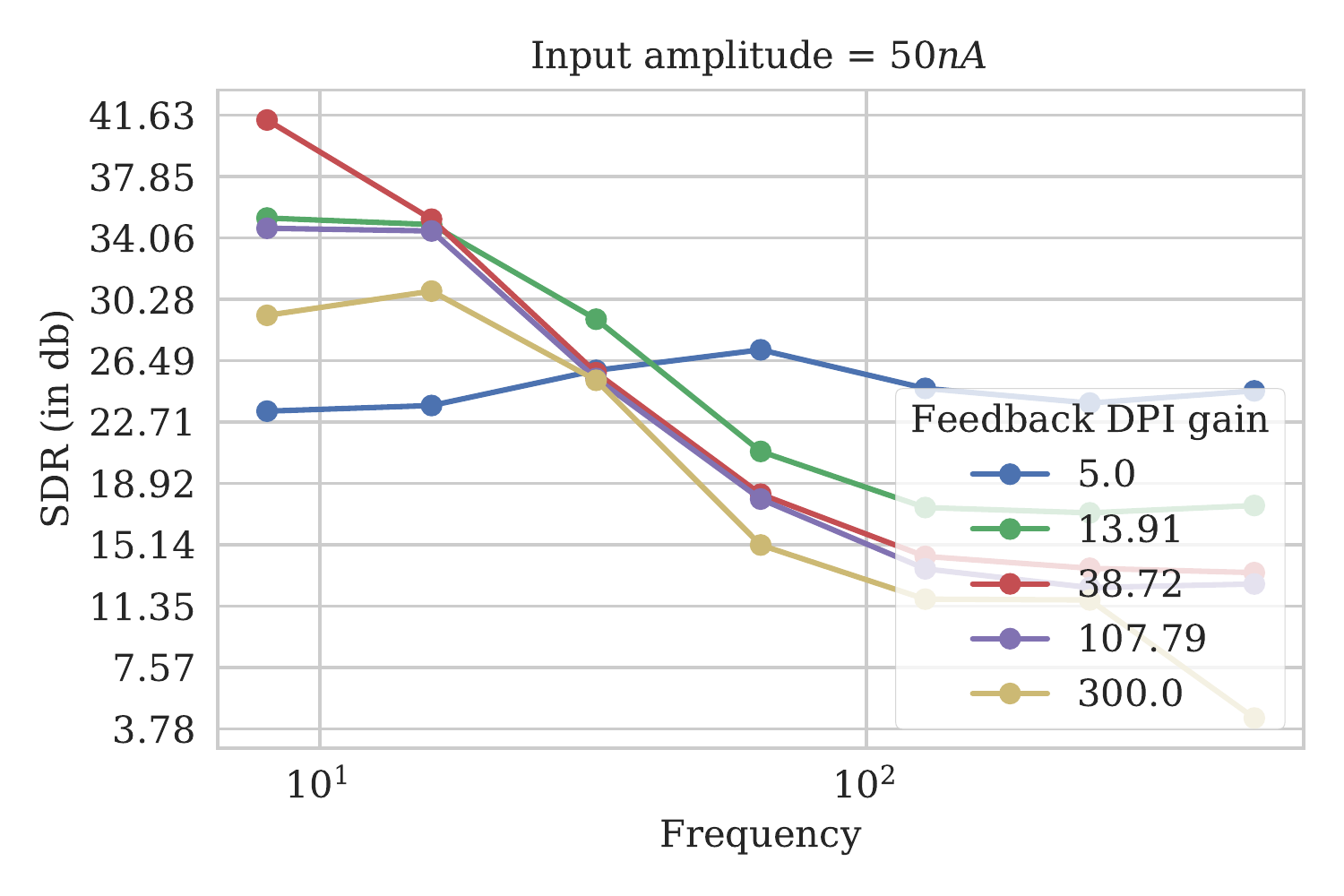}}
\subfloat[Energy consumed (\sd)\label{sd_fbgain_freq_power}]{\includegraphics[width=0.5\linewidth]{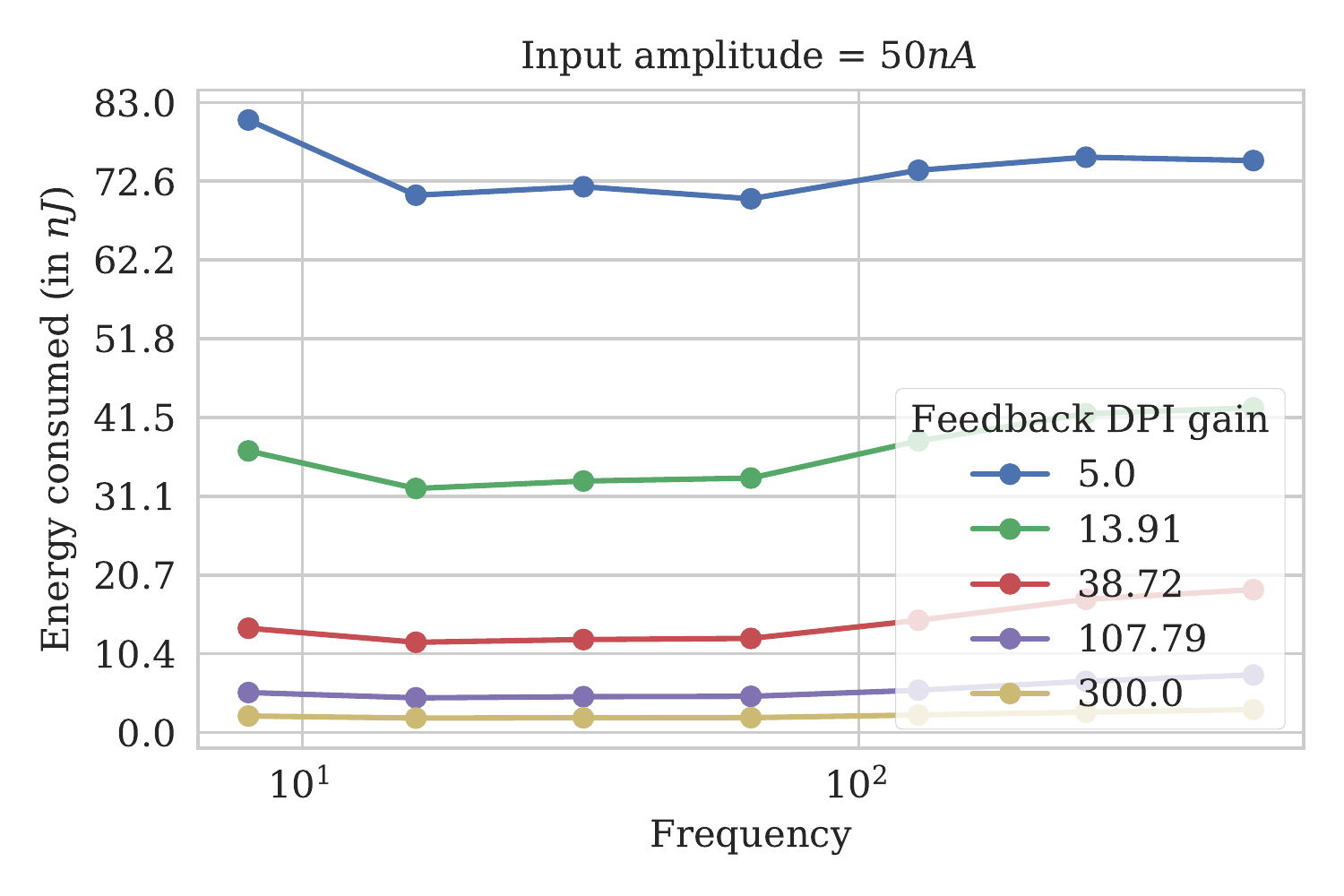}}\\
	\caption{Performance in 270 ms transient simulation.}
	\label{sdr}
\end{figure}

In this section, we compare the performance for two neuron circuits when transmitting a $50 nA$ sinusoid riding on a $50 nA$ DC bias. The bias is essential because the circuits only encode positive inputs. Both circuits were set up with same filter settings. We only show measurements of \ac{SDR} and power when sweeping the input frequency for different values of gain in the feedback filter as shown in Fig.~\ref{sdr}. While the reduced energy consumption ($2\times$ to $40\times$) of the \sd circuit is clearly visible in Fig.~\ref{rolls_fbgain_freq_power} and Fig.~\ref{sd_fbgain_freq_power}, the factor of improvement is smaller than in Fig.~\ref{fig:max_fire_energy}. This is because of the power consumption in the feedback \ac{DPI} circuit. We observe that the new \sd neuron circuit also achieves better \ac{SDR} in Fig.~\ref{rolls_fbgain_freq_snr} and Fig.~\ref{sd_fbgain_freq_snr}.

\subsection{Reservoir implementation example}
\begin{figure}[!ht]
	\centering
	\subfloat[Architecture of the simulated network\label{network}]{\includegraphics[width=0.4\linewidth, trim={1cm 0 1cm 1cm},clip]{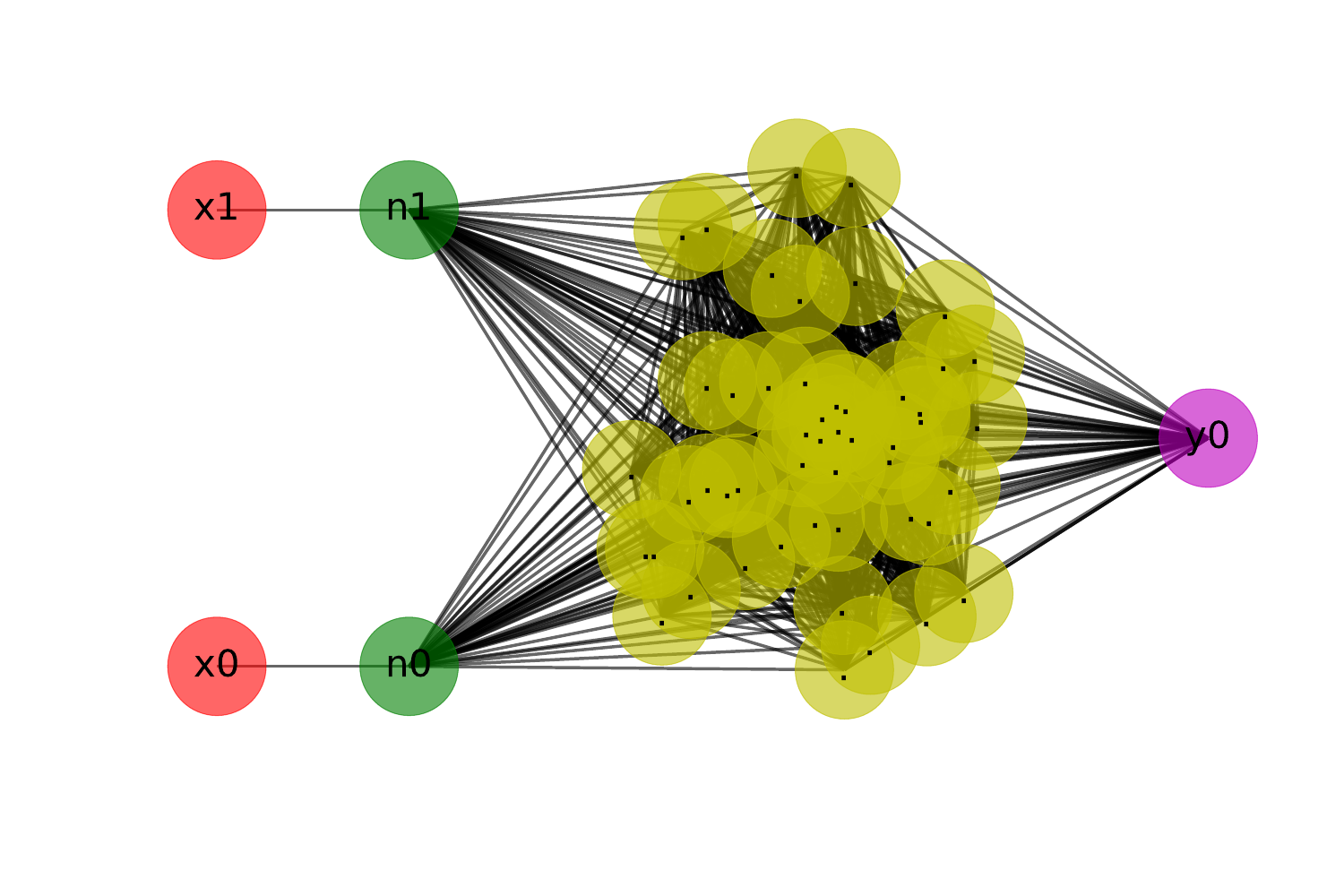}}
	\subfloat[Floating point ESN vs spiking ESN\label{res_ann}]{\includegraphics[width=0.6\linewidth]{ANN-SDNN}}\\
  \caption{Floating point \ac{ESN} mapped to a Spiking Neural Network (SNN)}
	\label{reservoir}
\end{figure}
A \sd neuron can be used to implement any high-resolution neural network models (ex. 32-bit floating-point) because the feedback signal, $s(t)$, is analogous to the state of the nodes in the network. This is because the event-driven communication model allows a spiking NN to behave like an analog system or at least as a digital system clocked at a very high frequency. (analogous to the equivalence between digital filters and their analog counterparts.)  Therefore, it is possible to map a high-resolution neural network model to an equivalent asynchronous spiking one by imposing some constraints. First, the bandwidth of the floating-point neurons should be suitably band-limited. This is not a limitation when interfacing the spiking neuromorphic systems to real-world sensors as they are naurally band-limited. Secondly, the inputs to the spiking system must be scaled such that spiking neurons do not saturate. This is easy to achieve in practice, by use of input gain correction. The mapping mechanism is especially useful to implement \ac{RNN}s where the improved ability of the \sd circuit to transmit temporally-changing signals is most beneficial. We show an example for this using an \ac{ESN} \cite{Jaeger02} with two inputs nodes (including bias), fifty recurrently connected nodes, and a single readout unit~(Fig.~\ref{network}). The bandwidth constraint is imposed by the retention factor, $0 < \alpha < 1$, in its state update equation:
\begin{align}
  s[n+1] = (1 - \alpha) s[n] + \alpha\sigma \left(x[n] W_{in} + s[n] W + b\right)
\label{eq:res}
\end{align}
where, $s[n]$ is the reservoir state in time step $n$, $x$ the input, $W_{in}$ the input connectivity matrix, $W$ the recurrent connectivity, $\sigma$ a non-linearity and $b$ the bias.
A transient simulation comparing the dynamics of the mapped and original networks is shown in Fig.~\ref{res_ann}. Source code is available online~\cite{Nair_19sdnn}.

\section{Comparison to other neuron implementations}
We only report area and power comparisons at a spike generation rate of 300 Hz as literature on temporal data encoding performance of spiking neurons is very scarce (see Table~\ref{tab:comparison}). The mixed-signal neuron designs in BrainScaleS\cite{Aamir_etal18BrainScales} and Neurogrid\cite{Benjamin_etal14} offer the closest comparison to this work. TrueNorth\cite{Merolla_etal14a}, Loihi\cite{Davies_etal18} and ODIN\cite{Frenkel_etal18} are digital systems that use advanced processes, time-multiplexing of neurons and low supply voltages. They are also designed for high input and output data rates and are arguably less suited than the proposed neuron design for low-bandwidth sensor data processing applications. Nevertheless, the proposed \sd neuron circuit consumes the lowest power of the surveyed implementations. Use of advanced processes and smaller supply voltages will further improve the performance of the presented neuron circuit. Finally, we would like to highlight that the presented circuit techniques (\sd model and regenerative feedback) are compatible with and beneficial for many of the neuromorphic systems proposed in literature.
\begin{table}[h!]
  \caption{Neuron performance comparison}
  \label{tab:comparison}
  \centering
\begin{tabular}{|l|l|l|l|l|}
\hline
\textbf{Project} & \textbf{Tech.} & \textbf{Supply} & \textbf{pJ/spike} & \textbf{Area/neuron} \\ \hline
BrainScaleS\cite{Aamir_etal18BrainScales}  & 65 nm  & 2.5/1.2 V  & 200 & 3372 $\mu m^2$ \\ \hline
Neurogrid \cite{Benjamin_etal14} & 180 nm & 3 V & 8000(est) & 1800 $\mu m^2$ \\ \hline
ODIN\cite{Frenkel_etal18} & 28 nm & 0.55-1.0V & 54 & 74 $\mu m^2$ (est) \\ \hline
TrueNorth\cite{Merolla_etal14a}  & 28nm & 0.7-1.05V  & 26(est) & $14.3 \mu m^2$ \\ \hline
Intel Loihi\cite{Davies_etal18} & 14 nm & 0.5-1.25V & 24 (est)  & $14 \mu m^2$ (est) \\ \hline
\textbf{This work} & \textbf{180nm} & \textbf{1.8V} & \textbf{10} & \textbf{2025 $\mu m^2$} \\ \hline
\end{tabular}
\end{table}

\section{Discussion and conclusion}
We presented a silicon neuron design implementing a \sd encoder and showed simulations for its targetted use cases. The equivalence to \sd encoders can be further improved by endowing the neuron with the ability to integrate negative errors, using a different log-domain filter design with lesser non-linear distortions, using more accurate circuits to compute the difference between the input and feedback signals. A key novelty of this paper is in the use of regenerative feedback to create a neuromorphic \sd circuit that can operate with orders of magnitude lower energy-consumption than existing neuron implementations, and in its application to real-valued \ac{RNN}s implemented with spiking neuromorphic systems.

\renewcommand{\bibfont}{\normalfont\small} % small biblio font
\printbibliography
% that's all folks

\end{document}